\documentclass{sig-alternate}
\usepackage{latexsym}
\usepackage{amsmath}

\def\punto{$\hspace*{\fill}\Box$}
\newcommand{\nop}[1]{}
\newcommand{\tuple}[1]{{\langle#1\rangle}}
\def\lBrack{\lbrack\!\lbrack}
\def\rBrack{\rbrack\!\rbrack}
\newcommand{\Bracks}[1]{\lBrack#1\rBrack}

\newtheorem{theorem}{Theorem}[section]

\newtheorem{example}[theorem]{Example}

\newtheorem{corollary}[theorem]{Corollary}

\usepackage{pst-tree}

\newcommand\Tsubtreel[1]{\pstree[levelsep=7.3mm]{\Tdot^{#1}}{\Tfan}}
\newcommand\Tsubtreer[1]{\pstree[levelsep=7.3mm]{\Tdot_{#1}}{\Tfan}}

\newcommand{\vcp}[3]{\ncarc[linestyle=dashed,arcangle=#3]{->}{#1}{#2}
                     \naput{copy}}
\newcommand{\vcpb}[3]{\ncarc[linestyle=dashed,arcangle=#3]{->}{#1}{#2}
                     \nbput{copy}}
\newcommand{\vcpl}[3]{\ncarc[linestyle=dashed,arcangle=#3]{<-}{#2}{#1}
                         \nbput{copy}}

\newcommand{\vcpmove}[3]{\ncarc[linestyle=dashed,arcangle=#3]{->}{#1}{#2}
                         \naput{move}}
\newcommand{\vcpmovel}[3]{\ncarc[linestyle=dashed,arcangle=#3]{<-}{#2}{#1}
                         \nbput{move}}

\newcommand{\Ttup}{\Tcircle{\tiny $\tuple{}$}}
\newcommand{\Ttupn}[1]{\Tcircle[name=#1]{\tiny $\tuple{}$}}
\def\Tset{\Tcircle{$*$}}
\newcommand{\Tsetn}[1]{\Tcircle[name=#1]{$*$}}

\newcommand{\TRtxt}[3]{\TR{{#1} {#3} {#2} {#3} {{\white #1}}}}
\newcommand{\TRtxtr}[2]{\TR{{\white #1} $\quad$ #2 $\quad$ {#1}}}
\newcommand{\TRtxtl}[2]{\TR{{#1} $\quad$ #2 $\quad$ {\white #1}}}

\newcommand{\vcptom}[1]{{\cal M}\Bracks{#1}}

\psset{treemode=D,levelsep=1.2cm}
\psset{radius=6pt,dotsize=4pt,treefit=loose,arrowinset=0,nrot=:U}

\title{A Visual Query Language for Complex-Value Databases}

\author{Christoph  Koch \\
Saarland University Database Group, Saarbr\"ucken, Germany \\
koch@infosys.uni-sb.de}

\date{}

\toappear{}

\begin{document}

\maketitle

\begin{abstract}
In this  paper, a visual  language, VCP, for queries  on complex-value
databases is proposed.  The main  strength of the new language is that
it is purely visual: (i) It has no notion of variable, quantification,
partiality, join,  pattern matching, regular expression, recursion,
or any other
construct  proper  to logical,  functional,  or  other database  query
languages and  (ii) has a  very natural, strong, and  intuitive design
metaphor.   The main operation  is that  of copying  and pasting  in a
schema tree.

We  show that  despite its  simplicity, VCP  precisely  captures complex-value
algebra  without  powerset, or  equivalently,  monad  algebra  with union  and
difference. Thus, its expressive power  is precisely that of the language that
is usually considered to play the role of relational algebra for complex-value
databases.
\end{abstract}

\section{Introduction}

Even though most  modern  database query  languages  are  based  on logical
or  algebraic
foundations (or a combination  of these, as is the case for  SQL) to allow for
the  declarative or  at least  abstract specification  of queries,
\nop{Users can
focus on {\em  what}\/ to obtain by their queries, rather  than on {\em how}\/
to obtain it.  In order to allow for efficient evaluation, query languages are
usually  restricted  in  expressive  power  (as  compared  to  Turing-complete
programming  languages).  This facilitates  the  automatic  mapping from  user
queries to  a good physical  query plan as  well as the optimization  of query
plans.   Restricting  the  expressiveness  of  query  languages  (while  still
retaining enough  power to  express the most  common queries) tends  to render
query languages easier  to learn and use.  This is  particularly true if query
languages are restricted to capture  precisely the queries expressible in some
well-founded framework such as  first-order logic or, equivalently, relational
algebra.
Nevertheless, it turns out that
} % end nop
inexperienced users are often overwhelmed
by the task of writing queries.
This has motivated efforts to develop
easy-to-use {\em visual}\/ query languages.

The  essential issue in  defining good  visual query  languages is  to provide
strong visual metaphors for the constructs in a query and the steps to be taken
to  define it.  Only by  such strong  visual metaphors  can a  visual language
become easy to use for inexperienced users. 
Examples of such metaphors for data management operations
in the wider sense are, e.g., deletion by dragging an icon representing a data
object onto a garbage can or data relocation by dragging and dropping icons in
a directory tree.

We distinguish between (1) graphical  (or graph-based) query languages such as
QBE \cite{QBE}  and Graphlog \cite{CoMe90}  and (2) visual languages  in which
the specification of the query is a process, i.e., the query is defined by the
interaction   with   the   user   rather   than  by   the   static   graphical
outcome.\footnote{Languages of the first  class are traditionally  called
visual as  well.}  For  example, manipulating a  directory hierarchy in  an
operating system
window manager  usually involves a  sequence of insertion, copy,  and deletion
steps  which together  define the transformation on  the file
system  to  be performed.  By  just considering  the  outcome  -- a  directory
structure -- it is not possible to tell which transformation was carried out.

The  interaction of  a user  with a  system may  give a  surprising  amount of
expressive   power   while   allowing    the   visual   approach   to   remain
intuitive. Static  graphical languages tend to offer  more powerful constructs
(such  as variables and  quantifiers) to  compensate for  this.  Even  if such
constructs  are   provided  via  graphical  objects,   queries  still  closely
correspond to traditional  textual query languages, and much  of the appeal of
visual specification is lost.

QBE  \cite{QBE} is a  visual query  language for  relational databases  and an
attractive alternative  to relational algebra.  In its original
form, it is  widely taught but has  failed to have a large  impact on database
practice.  One  reason for this  may be its  reliance on variables,  which are
required  to perform  joins,  and  may be  a  concept hard  to  deal with  for
non-expert  users.  One  seemingly close  cousin of  QBE, the  visual language
employed in Microsoft Access, avoids the use of variables by replacing them by
lines that connect table columns to be joined.  Still, lines are an unsuitable
metaphor  for multi-argument  and  multi-way joins.   The otherwise  intuitive
design metaphor of QBE (and equally  the visual query language of Access) also
breaks when  one wants to  go beyond conjunctive  queries to employ  union and
difference in queries.

Even greater  problems are  encountered when a  visual language is  sought for
nested relational or complex-value data\-bases \cite{JS82,AB1988,AHV95}.
QBE-like tables
can be  visually nested  within each other  easily, but  there seems to  be no
clear -- and  sufficiently expressive -- semantics to  multiple occurrences of
variables (which  give us joins  in QBE).  Such  a semantics would need  to be
able to  express functionality beyond joins  such as nesting  and unnesting as
well.  Previous  work on languages  for complex-value databases  has therefore
either  resulted in  expressively  rather weak  languages  or has  compromised
simplicity (the  existence of a strong  visual design metaphor)  to obtain the
right expressiveness.

Besides QBE,  another example  of a  language of the  first kind  is visXCerpt
\cite{BBSW2003},  a GUI-based  query  language for  XML that uses  graphical
primitives for notions such as variables, quantification, recursion,
and partiality.
The expressive po\-wer of visXCerpt has not been formally studied, but appears
to be very high (Turing-complete).
QURSED \cite{PPV2002}, another GUI-based XML query language, trades in
expressiveness for simplicity (even though no formal study of its expressive
power is available). There are no explicit graphical objects for
variables, but the user is required to have a notion of concepts such as
variables and Boolean combinations of conditions in order to successfully use
the query builder.
Both the visXCerpt and the QURSED tool use a number of visual
operations to build queries (such as dragging and dropping data locations);
however, the query is the static outcome of this process.

XQBE \cite{XQBE}  and XML-GL \cite{XMLGL} are graphical  languages for queries
on XML. In  both, there is no notion of explicit  variables; instead lines are
drawn similarly to the visual tool of MS Access.  A query consists of a source-
and  a construct-part  (both based  on graphical  schema  representations). No
study of the expressive power of these languages is available, but the queries
seem to correspond to the nested (XML) analogs of
restricted classes of conjunctive queries. 

In Graphlog \cite{CoMe90}, queries are  graphs with node and edge annotations,
which are  to be matched  against a graph  database (in the spirit  of finding
subgraph homomorphisms).  Graphlog handles a class of linear recursive
queries by  allowing for  certain regular expressions  over relation  names on
edges, where such  relations can be defined by Graphlog  query graphs and used
cyclically.  When recursion is used sparingly, Graphlog queries tend to be
easy to read.  The expressive power of Graphlog
is studied formally in \cite{CoMe90} and it
turns out that the language  has a number of nice characterizations.  However,
only binary relations  can be defined, so no  general data transformations are
possible and the expressiveness does not match that required for complex-value
databases.
Similar approaches are taken in G-Log \cite{PPT1995} and in GraphLog's
predecessor G+ \cite{CMW1988}.

Lixto \cite{BFG2001a},  a visual Web wrapper  generator, is based  on an
intuitive  metaphor  for  information  extraction from  Web  pages.  (The
selection  of regions in  a Web  page with  the mouse  and the  association of
``patterns''  to  such  regions.)   However,  Lixto only  covers  unary  (tree
node-selecting) que\-ries rather than data transformation queries, which
is sufficient in the wrapping context. 
The core language of Lixto has been shown to capture
precisely an important and well-studied
class of queries, those definable in monadic
second-order logic  over trees \cite{GKJACM}; however, to  achieve this, Lixto
has to resort
to  recursion  \cite{BFG2001b},  for  which  no  visual  metaphor  is
provided.

To  this   day,  there  is  no   truly  visual  language   that  captures  the
expressiveness of  any of the clean  theoretically-founded data transformation
languages such  as relational  algebra or complex-value  algebra
\cite{AB1988,AHV95}.
This paper  aims to  improve on  this situation by  proposing a  language that
appears to satisfy these desiderata.
Our contributions are as follows.
\begin{itemize}
\item
A visual  language, VCP, for queries  on complex-value
databases is proposed.  The main  strength of the new language is that
it is purely visual: (i) It has no notion of variable, quantification,
partiality, join,  pattern matching, regular expression, recursion,
or any other
construct  proper  to logical,  functional,  or  other database  query
languages and  (ii) has a  very natural, strong, and  intuitive design
metaphor.

The  only operations  available and  needed  in VCP  are that  of copying  and
pasting in a {\em schema tree}\/  as well as inserting, renaming, and deleting
nodes and filtering sets.
A  schema tree only  uses three kinds  of nodes, namely  set-typed and
tuple-typed nodes and atomic value leaves.

The only (oblique) advanced
notion that the  user has to deal with is  that of a
collection (a set). Schema trees  can be understood and treated similarly
to directory trees as they have become commonplace in OS window
managers (with directories as collections of files).

\item
We  show  that despite  its  simplicity,  VCP  precisely captures  the
complex-value   algebra without powerset \cite{AB1988}, or equivalently,
monad algebra with union and difference \cite{TBW1992,BNTW1995}. Thus, its
expressive power  is precisely  that of the  language that  is usually
considered to  play the role  of relational algebra  for complex-value
databases (cf.\ \cite{AB1988,AHV95}).

These languages are also a foundation of -- and probably
very similar in expressive power too -- to XML query languages such as
XQuery. Even though this remains to be verified, this renders it likely
that VCP may give rise to the very first well-founded visual XML
query language.
\end{itemize}

VCP is a member of the second class of query languages defined above --
queries are defined as an interactive process. It turns out that this
interaction gives us so much expressiveness that we need only very simple
query constructs and operations.

\begin{figure*}
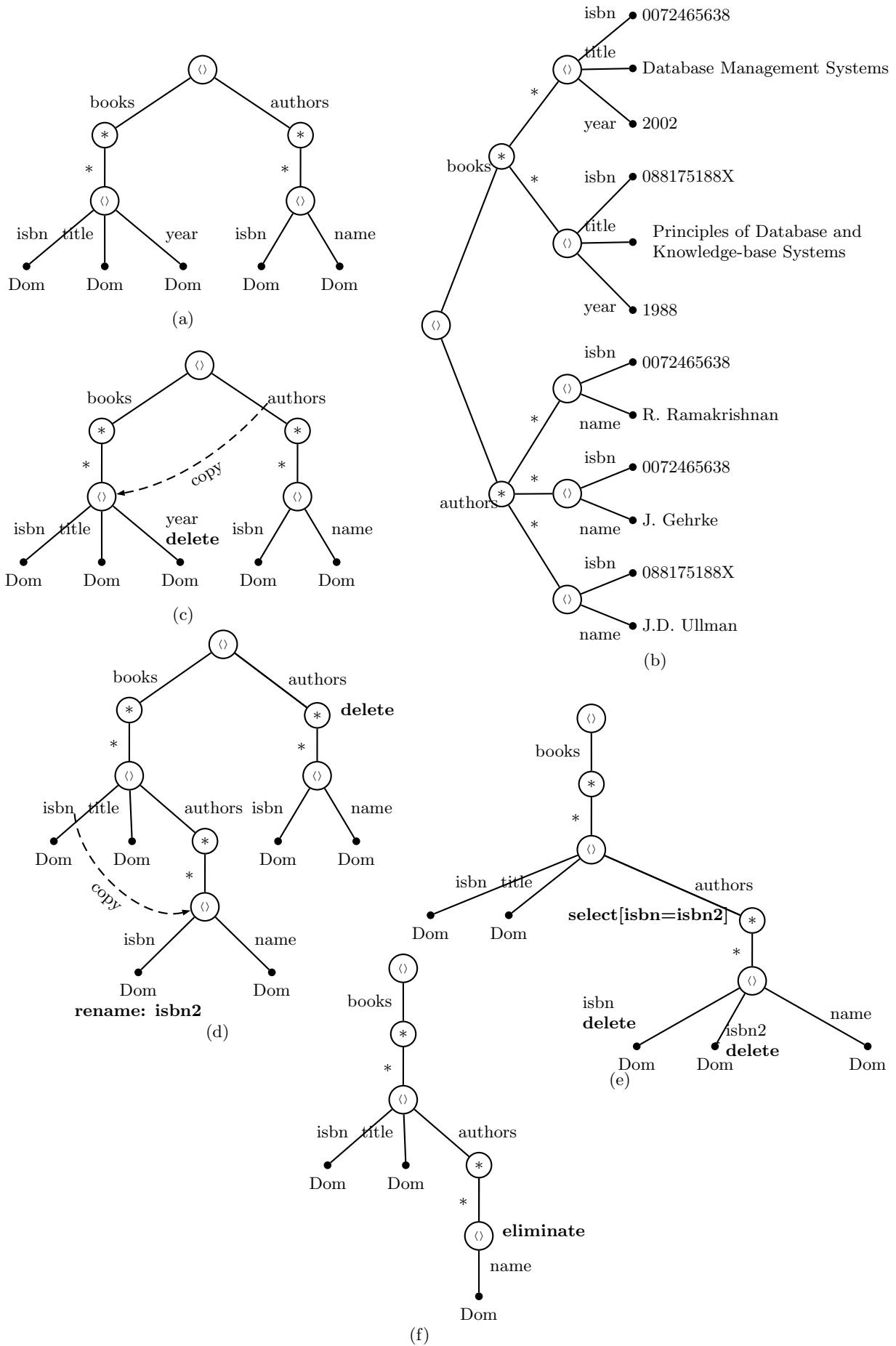

\begin{center}
\psset{treefit=tight}
\vspace{-0.5cm}
\begin{tabular}{c}
\pstree{\Ttup}
{
   \pstree{\Tset^{books}}
   {
      \pstree{\Ttup^{$*$}}
      {
         \Tdot~{Dom}^{isbn}
         \Tdot~{Dom}^{title}
         \Tdot~{Dom}_{year}
      }
   }
   \pstree{\Tset_{authors}}
   {
      \pstree{\Ttup^{$*$}}
      {
         \Tdot~{Dom}^{isbn}
         \Tdot~{Dom}_{name}
      }
   }
}
\\
\\
(a)
\\
\\
\pstree{\Ttup}
{
   \pstree{\Tset^{books}}
   {
      \pstree{\Ttupn{To}^{$*$}}
      {
         \Tdot~{Dom}^{isbn}
         \Tdot~{Dom}^{title}
         \Tdot~{Dom}_{\hspace{-4mm}\begin{tabular}{l}year \\ {\bf delete}\end{tabular}}
 %        \Tdot~{Dom}_{publisher}
      }
   }
   \pstree{\Tset_{\pnode{From}authors}}
   {
      \pstree{\Ttup^{$*$}}
      {
         \Tdot~{Dom}^{isbn}
         \Tdot~{Dom}_{name}
      }
   }
}
\vcpl{From}{To}{-20}
\\
\\
(c)
\end{tabular}
\hspace{5mm}
\begin{tabular}{c}
\pstree[treemode=R]{\Ttup}
{
   \pstree{\Tset^{books}}
   {
      \pstree{\Ttup^{$*$}}
      {
         \Tdot~{0072465638}^{isbn}
         \Tdot~{Database Management Systems}^{title}
         \Tdot~{2002}_{year}
      }
      \pstree{\Ttup^{$*$}}
      {
         \Tdot~{088175188X}^{isbn}
         \Tdot~{\begin{tabular}{l}
Principles of Database and \\ Knowledge-base Systems\end{tabular}}^{title}
         \Tdot~{1988}_{year}
      }
   }
   \pstree{\Tset_{authors}}
   {
      \pstree{\Ttup^{$*$}}
      {
         \Tdot~{0072465638}^{isbn}
         \Tdot~{R.~Ramakrishnan}_{name}
      }
      \pstree{\Ttup^{$*$}}
      {
         \Tdot~{0072465638}^{isbn}
         \Tdot~{J.~Gehrke}_{name}
      }
      \pstree{\Ttup^{$*$}}
      {
         \Tdot~{088175188X}^{isbn}
         \Tdot~{J.D.~Ullman}_{name}
      }
   }
}
\\
\\
(b)
\end{tabular}

\vspace{-8mm}

\hspace{-8.5cm}
\pstree{\Ttup}
{
   \pstree{\Tset^{books}}
   {
      \pstree{\Ttup^{$*$}}
      {
         \Tdot~{Dom}^{isbn\pnode{From}}
         \Tdot~{Dom}^{title}
         \pstree{\Tset_{authors}}
         {
            \pstree{\Ttupn{To}^{$*$}}
            {
               \Tdot~{\begin{tabular}{c}Dom \\
{\bf rename: isbn2}
\end{tabular}}^{isbn}
               \Tdot~{Dom}_{name}
            }
         }
      }
   }
   \pstree{\TRtxtr{{\bf delete}}{\Tset}_{authors}}
   {
      \pstree{\Ttup^{$*$}}
      {
         \Tdot~{Dom}^{isbn}
         \Tdot~{Dom}_{name}
      }
   }
}
\vcpb{From}{To}{-50}

\hspace{-8.5cm}
(d)

\vspace{-6.2cm}

{~}~\hspace{8cm}
\pstree{\Ttup}
{
   \pstree{\Tset^{books}}
   {
      \pstree[treefit=tight]{\Ttup^{$*$}}
      {
         \Tdot~{Dom}^{isbn\pnode{From}}
         \Tdot~{Dom}^{title}
         \pstree{\TRtxtl{{\bf select[isbn=isbn2]}}{\Tset}_{authors}}
         {
            \pstree{\Ttupn{To}^{$*$}}
            {
               \Tdot~{Dom}^{~\hspace{4mm}\begin{tabular}{l}
isbn \\ {\bf delete}\end{tabular}}
               \Tdot~{Dom}-{\hspace{-1mm}\begin{tabular}{l}
isbn2 \\ {\bf delete}\end{tabular}}
               \Tdot~{Dom}_{name}
            }
         }
      }
   }
}

{~}~\hspace{6cm}
(e)

\vspace{-2.5cm}

%\hspace{-1cm}
\pstree{\Ttup}
{
   \pstree{\Tset^{books}}
   {
      \pstree{\Ttup^{$*$}}
      {
         \Tdot~{Dom}^{isbn}
         \Tdot~{Dom}^{title}
         \pstree{\Tset_{authors}}
         {
            \pstree{\TRtxtr{{\bf eliminate}}{\Ttup}^{$*$}}
            {
               \Tdot~{Dom}_{name}
            }
         }
      }
   }
}

\vspace{1mm}

\hspace{-1cm}(f)
\end{center}
\vspace{-5mm}
\caption{Initial data tree (b) and query steps (a), (c)-(f) of
Example~\ref{ex:1}.}
\label{fig:ex1a}
\end{figure*}

\begin{figure}[t]
\begin{center}
%{~}~\hspace{-8cm}
\pstree{\Ttup}
{
   \pstree{\Tset^{books}}
   {
      \pstree{\Ttup^{$*$}}
      {
         \Tdot~{Dom}^{isbn\pnode{From}}
         \Tdot~{Dom}^{title}
         \pstree{\Tset_{authors}}
         {
               \Tdot~{Dom}_{$*$}
         }
      }
   }
}

%{~}~\hspace{-8cm}
(a)

%\vspace{-5cm}

\vspace{5mm}

%{~}~\hspace{8cm}
\pstree[treemode=R,treefit=loose]{\Ttup}
{
   \pstree{\Tset^{books}}
   {
      \pstree{\Ttup^{$*$}}
      {
         \Tdot~{0072465638}^{isbn}
         \Tdot~{Database Management Systems}_{title}

   \pstree{\Tset_{authors}}
   {
      \Tdot~{R.~Ramakrishnan}^{$*$}
      \Tdot~{J.~Gehrke}_{$*$}
   }

      }
      \pstree{\Ttup^{$*$}}
      {
         \Tdot~{088175188X}^{isbn}
         \Tdot~{\begin{tabular}{l}
Principles of Database and \\ Knowledge-base Systems\end{tabular}}_{title}

   \pstree{\Tset_{authors}}
   {
      \Tdot~{J.D.~Ullman}_{$*$}
   }

      }
   }
}

\vspace{3mm}

%{~}~\hspace{8cm}
(b)
\end{center}
\vspace{-5mm}
\caption{Conclusion of Example~\ref{ex:1}.}
\label{fig:ex1b}
\end{figure}

\begin{example}
\label{ex:1}
\em
We discuss an example VCP query informally to provide a first idea of the
language. Consider a relational database with two relations
$books(\underline{isbn}, title, year)$ and $authors(\underline{isbn, name})$,
modeling books and their (possibly multiple) authors.

Throughout this paper, we will focus on a complex-value data model in which
relations may be nested into each other. A graphical (tree-based)
representation of the schema (a {\em schema tree}\/) is shown in
Figure~\ref{fig:ex1a}~(a),
and detailed definitions will follow in the technical sections of
the paper. ``Dom'' denotes the domain of atomic values such as strings and
integers.
A corresponding data tree -- modeling two books and their authors -- is
shown in Figure~\ref{fig:ex1a}~(b).

We will specify a query that maps every complex value
database of our given schema
to a complex value consisting of a set of book-tuples of the form
$
\tuple{isbn, title, authors},
$
where ``authors'' is the set of authors of the book. That is, this query
{\em nests}\/ the authors of each book into their book tuple.
The intended query result and a corresponding schema tree can be found
in Figures~\ref{fig:ex1b}~(b) and (a), respectively.

In VCP, we can define this query visually, by a number of
interactive modifications of the schema tree. We proceed as follows.
(1) First we execute a bulk copy operation of the ``authors'' relation into
each tuple of the ``books'' relation by simply dragging the ``authors'' subtree
of the schema tree onto the node representing the ``books'' tuples
(adding another relation-typed attribute/column to the ``books'' relation).
This is shown in Figure~\ref{fig:ex1a}~(c).

(2) Then we delete the ``year'' edge from the schema tree, which
removes this attribute from ``books'' and thus the years from all book
entries in the database. (Because of space limitations,
not all steps have their own figure; this operation can be found in
Figure~\ref{fig:ex1a}~(c).)

(3) Next we make a bulk copy of the ``isbn'' attribute of each book
tuple $t$ to each of the author tuples in the ``authors'' attribute of $t$.
This is performed by dragging the isbn edge emanating from the node
representing the book tuples to the node representing the author tuples
nested inside books (see Figure~\ref{fig:ex1a}~(d)). In order not to
have two ``isbn'' attributes in the nested authors, though, we first
rename the ``isbn'' attribute of the nested authors to ``isbn2''.

(4) We will not need the original authors relation anymore, so we can
remove it from the complex value computed as the query result by
deleting the top-level authors subtree from the schema tree
(Figure~\ref{fig:ex1a}~(d)).

(5) Now we apply a ``select'' operation on the schema tree node
corresponding to the ``authors'' relation nested inside the books and filter
out those authors for which isbn $\neq$ isbn2, i.e., for each book tuple
$t$, we remove those
author tuples that really belong to book $t$ (Figure~\ref{fig:ex1a}~(e)).

(6) Then we can eliminate the ``isbn'' and ``isbn2'' attributes of the
nested author tuples by deleting their subtrees
(Figure~\ref{fig:ex1a}~(e)).

(7) Finally, we eliminate the ``authors'' tuple node -- which has only one
child (in other terms, a single attribute).
This transforms, for each book tuple,
the ``authors'' attribute from a set of unary tuples to a set of domain
values (author names) (Figure~\ref{fig:ex1a}~(e)).
\footnote{Visually,
this operation cuts out a node from a path in the schema tree,
rather than deleting the subtree rooted by the node eliminated.}

As stated above, the final schema tree and the query result are shown in
Figure~\ref{fig:ex1b}.
\punto
\end{example}

Further examples  of VCP queries can be found in 
Figures~\ref{fig:cartprod}, \ref{fig:minus}, and \ref{fig:nest}.
However, these examples fulfill a double duty, and serve to demonstrate
certain expressiveness arguments. They are somewhat more abstract.

The structure of the paper is as follows. First, in
Section~\ref{sect:preliminaries}, we introduce types for complex values and
their corresponding schema trees. Moreover, we give an introduction to
monad algebra. Section~\ref{sect:vcp} presents the language VCP and gives
a number of examples. In Section~\ref{sect:expressiveness}, it is shown that
VCP precisely captures the expressive power of monad algebra.
We conclude with a discussion of VCP and future work
(Section~\ref{sect:discussion}).

\section{Preliminaries}
\label{sect:preliminaries}

\subsection{Schema Trees}

We model complex values constructed from sets, tuples, and atomic values
from a single-sorted domain\footnote{All results in this paper immediately
generalize to many-sorted domains.} in the normal way. Types are terms
of the grammar
\[
\tau ::= \mbox{Dom} \mid \{ \tau \}
\mid \tuple{A_1: \tau_1, \dots, A_k: \tau_k}
\]
where $k \ge 0$.
Type terms have an obvious tree representation; we will call such trees
{\em schema trees}\/.

Note that  schema trees have  three kinds of  nodes -- tuple, set,  and atomic
value type nodes  -- and two kinds  of edges -- tuple and  set-edges, of which
the tuple edges carry an attribute name  as label. An example of a schema tree
for     the      type     $\tuple{R:\{\tuple{A:\tau_1,     B:\tau_2}\},     S:
  \{\tuple{C:\tau_3,D:\tau_4}\}}$        can         be        found        in
Figure~\ref{fig:cartprod}~(a).   If  $\tau_1=\tau_2=\tau_3=\tau_4=\mbox{Dom}$,
this  type  is an  appropriate  representation  of  relational schema  $R(AB),
S(CD)$.
(Here and later we model a relational data\-base
as a {\em tuple}\/ of relations to get a single complex value.)

\subsection{Monad Algebra}

\nop{
Structural recursion on union representation of sets:

\begin{tabular}{lrl}
fun & $\mbox{flatmap}(f)(\{\})$ =& \{\} \\
$\mid$ & $\mbox{flatmap}(f)(\{ S \})$ =& $f(S)$ \\
$\mid$ & $\mbox{flatmap}(f)(S_1 \cup S_2)$ =&
    $\mbox{flatmap}(f)(S_1) \cup \mbox{flatmap}(f)(S_2)$
\end{tabular}

($\mbox{flatmap}(f) = \mbox{map}(f) \circ \mbox{flatten}$)

{\bf Monad algebra} \cite{TBW1992}:
flatmap + some very minor lambda abstraction to create pairs and
take elements of them.

This is already quite powerful:
\begin{eqnarray*}
\mbox{flatten} &:=& \mbox{flatmap}(\lambda S.S), \mbox{ type } \{\{ \tau \}\} \rightarrow \{ \tau \} \\
\mbox{map} &:=& \mbox{flatmap}(\lambda x.\{ f x \}) \\
&& (\mbox{If $f: \sigma \rightarrow \tau$ then map($f$): $\{ \sigma \} \rightarrow \{ \tau \}$}) \\
\mbox{pairwith}_2(x, S) &:=& \mbox{map}(\lambda y.(x,y))S
\end{eqnarray*}
} % end nop

Consider the query language on complex values
consisting of expressions built from the
following operations (the types of the operations are provided as well):
\begin{enumerate}
\item
identity
\[
\mbox{id}: x \mapsto x \quad\quad \tau \rightarrow \tau
\]

\item
composition
\[
f \circ g: x \mapsto g(f(x))
\quad\quad
\frac{f: \tau \rightarrow \tau',\; g: \tau' \rightarrow \tau''}
{f \circ g: \tau \rightarrow \tau''}
\]

\item
constants from $\mbox{Dom} \cup \{ \emptyset, \tuple{} \}$
($\tuple{}$ is the nullary tuple)

\item singleton set construction
\[
\mbox{sing}: x \mapsto \{ x \}
\quad\quad
\tau \rightarrow \tau'
\]

\item application of a function to every member of a set
\[
\mbox{map}(f): X \mapsto \{ f(x) \mid x \in X \}
\]
\[
\frac{f: \tau \rightarrow \tau'}{\mbox{map}(f): \{ \tau \} \rightarrow
\{ \tau' \}}
\]

\item flatten:
$X \mapsto \bigcup X \quad\quad \{\{\tau\}\} \rightarrow \{ \tau \}$

\item pairing\footnote{Operations pairwith$_{A_i}$ can be defined analogously.}
\begin{multline*}
\mbox{pairwith}_{A_1}:
\langle A_1: X_1, A_2: x_2, \dots, A_n: x_n \rangle \mapsto \\
 \{ \langle A_1: x_1, A_2: x_2, \dots, A_n: x_n \rangle \mid x_1 \in X_1 \}
\end{multline*}
\begin{multline*}
\tuple{A_1: \{ \tau_1 \}, A_2: \tau_2, \dots, A_n: \tau_n } \rightarrow \\
\{ \tuple{A_1: \tau_1, \dots, A_n: \tau_n} \}
\end{multline*}

\item
tuple formation
\begin{multline*}
\langle A_1: f_1, \dots, A_n: f_n \rangle: \\
%   { }~{ } \quad\quad
   x \mapsto \langle A_1: f_1(x), \dots, A_n: f_n(x) \rangle
\end{multline*}
\[
\frac{f_1:\tau \rightarrow \tau_1, \dots, f_n:\tau \rightarrow \tau_n}
{\langle A_1: f_1, \dots, A_n: f_n \rangle: \tau \rightarrow
\tuple{A_1: \tau_1, \dots, A_n: \tau_n}}
\]

\item
projection
\[
\pi_{A_i}: \langle A_1: x_1, \dots, A_i: x_i, \dots, A_n: x_n \rangle
   \mapsto x_i
\]
\[
\pi_{A_i}: \tuple{A_1: \tau_1, \dots, A_n: \tau_n} \rightarrow \tau_i
\]
\end{enumerate}

The language has a strong and clean theoretical foundation from programming
language theory, for which we have to refer to e.g.\ \cite{TBW1992}.

\nop{
The language has a nice theoretical foundation from programming
language theory, that of structural recursion on sets extended by a small
amount of machinery for creating and destroying tuples \cite{TBW1992}.
Formally, the language above is a Cartesian category with a
{\em strong monad}\/ on it (where ``strong'' refers to
so-called {\em tensorial strength}\/
introduced by the ``pairwith'' operation).
We thus call this language  {\em monad algebra}\/ \cite{TBW1992}, or
${\cal M}$ for short.
By this observation, we can get an entire equational theory for this query
language -- which can be used for query optimization -- for free, to be
taken from books on category theory (e.g.\ \cite{MacLane}).
} % end nop

Returning to the definition of our operations,
note that projection is applied to tuples rather than to sets of tuples
as in relational algebra. For example, the relational algebra expression
$\pi_{AB}$ (on some relation with at least columns $A$ and $B$)
corresponds to $\mbox{map}(\tuple{A:\pi_A, B:\pi_B})$ in ${\cal M}$.

\begin{table*}[th]
\begin{small}
\begin{center}
\begin{tabular}{|l||l|l|p{5cm}|p{4.6cm}|l|}
\hline
& Name \& arguments & Applied to & Conditions & Function & Example \\
\hline
\hline
1 &
new constant ($A, c$) &
tuple node $v$ &
- &
Adds a new attribute $A$ with constant value
$c \in \mbox{Dom} \cup \{ \emptyset, \tuple{} \}$ to $v$. &
Fig.~\ref{fig:minus}~(d) \\
\hline
2 &
insert tuple ($A$) & node $v$ & - &
Replaces the subtree rooted by $v$ by $\tuple{A: v}$,
i.e., by a new tuple-node with $v$ as $A$-child. &
Fig.~\ref{fig:dupl}~(a) \\
\hline
3 &
insert set &
node $v$ &
- &
Replaces the subtree rooted by
$v$ by $\{ v \}$, i.e.\
by a new set-node with $v$ as child & \\
\hline
4 &
rename ($A$) &
tuple edge $e$ &
- &
Renames the label of $e$ (= an attribute of a tuple) to $A$ &
Fig.~\ref{fig:minus}~(b) \\
\hline
5 &
eliminate &
set node $v$ &
The parent node of $v$ is also a set node &
cuts $v$: let $v = \{ \tau \}$. Then, $v$ is replaced in the schema tree
by $\tau$.
% Flattens the set of sets into a set
&
Fig.~\ref{fig:cartprod}~(e) \\
\hline
6 &
eliminate &
tuple node $v$ &
The subtree rooted by $v$ is of the form
$\tuple{A: \tau}$, i.e., it has precisely one attribute
&
Replaces $v$ by $\tau$.
% i.e., performs the projection $\pi_A$
%
%if there are several children and the parent is a set, this could
%mean union. (bad visual intuition)
&
Fig.~\ref{fig:cartprod}~(d) \\
\hline
7 &
delete [subtree] &
tuple edge $e$ &
- &
Deletes $e$ and its subtree; reduces arity of tuple by one &
Fig.~\ref{fig:minus}~(f) \\
%
%\hline
%delete [subtree] &
%set edge &
%- &
%empties set (optional) & \\
%
\hline
8 &
copy to ($v$) &
tuple edge $e$ &
The label of $e$ must not exist yet in the
destination tuple node $v$.
The path from $\mbox{nca}(v,e)$ to $e$ must be $*$-free.
&
Adds $e$ and its subtree to tuple node $v$.
&
Fig.~\ref{fig:minus}~(c) \\
\hline
\hline
9 &
copy to ($v$) &
set edge $e$ &
The types below $e$ and destination set node $v$ must be equal.
The path from $\mbox{nca}(v,e)$
to $e$ must contain precisely one $*$ (i.e., the label of
$e$).
&
Does not modify schema tree.
% union-in
& \\
\hline
10 &
select ($A$, $B$) &
set node $v$ &
The child of $v$ must be a tuple
of type $\tuple{A:\tau_A, B:\tau_B, \dots}$ &
Does not modify schema tree.
% but selects those tuples of the set node $v$
% for which $A=B$ (deep equality)
&
Fig.~\ref{fig:minus}~(e) \\
\hline
%
%SetNode.subtract(SetEdge fromEdge)
\end{tabular}
\end{center}
\end{small}
\vspace{-3mm}
\caption{VCP operations.}
\label{tab:operations}
\end{table*}

By {\em positive monad algebra}\/
${\cal M}_\cup$, we denote ${\cal M}$ extended by the set union operation.
This language has a number of nice properties \cite{TBW1992, BNTW1995},
but it is known that it is incomplete as a practical query language because
it cannot yet express selection, set difference, or set intersection.

However, if we extend ${\cal M}_\cup$ by any nonempty subset of the operations
selection (of the form $\sigma_{A=B}$ on complex values of type
$\{ \tuple{A:\tau, B:\tau', \dots} \}$, where ``$=$'' denotes
deep equality of complex values),
set difference ``$-$'', set intersection $\cap$, or nesting\footnote{The
``nest'' operation of complex value algebra without powerset \cite{AB1988}
groups tuples by some of their attributes. For example,
$\mbox{nest}_{C=(B)}(R)$ on relation $R(AB)$ computes the value
$
\{ \tuple{A\!: x, C\!: \{ \tuple{B\!:y} \mid \tuple{A\!:x, B\!:y} \in R \}}
\mid (\exists y) \tuple{A\!:x, B\!:y} \in R \}.
$},
we always get the
same expressive power.
%
%\footnote{No analogous statement can be made about flat relational algebra.}
%
We will call any one of these extended languages {\em full monad algebra}.

\begin{theorem}[\cite{TBW1992}]
\label{theo:ma_equivalences}
${\cal M}_\cup[\sigma]$ = ${\cal M}_\cup[-]$ = ${\cal M}_\cup[\cap]$ =
%
%${\cal M}_\cup[\subseteq]$ =
%${\cal M}_\cup[\in]$ = ${\cal M}_\cup[=]$ =
%
${\cal M}_\cup[\mbox{nest}]$.
\end{theorem}

Moreover, generalizing selections to test against constants or to support
``$\in$'', ``$\subseteq$'', or Boolean combinations of conditions does not
increase the expressiveness of full monad algebra \cite{TBW1992}.

\begin{example}
\label{ex:m_minus}
{\em
The query of Example~\ref{ex:1} can be phrased in ${\cal  M}_\cup[\sigma]$
as
\begin{multline*}
\langle books: \mbox{pairwith}_{books} \circ \\
\mbox{map}(\langle isbn: \pi_{books} \circ \pi_{isbn},
   title: \pi_{books} \circ \pi_{title}, \\
   authors: \tuple{1: \pi_{books} \circ \pi_{isbn},
                   2: \pi_{authors}} \circ \\
\mbox{pairwith}_{authors} \circ
\mbox{map}(\langle isbn: \pi_1, \\
   isbn2: \pi_2 \circ \pi_{isbn},
   name: \pi_2 \circ \pi_{name} \rangle) \circ \\
\sigma_{isbn = isbn2} \circ \mbox{map}(\pi_{name})
\rangle ) \rangle
\end{multline*}

That is, we first pair each book tuple with the set of all authors.
Using the ``map'' operation, we then process each of these pairs as follows.
For each book-author pair, we pair the book's isbn number with each of the
author tuples. Then we are able to select those authors that belong
to the book (using $\sigma$). Each book tuple is made a triple of
an isbn number, a title, and a set of authors. The latter is a set of
names -- rather than tuples of isbn, isbn2, and name -- created by
mapping $\pi_{name}$ onto the set of author tuples.
}
\punto
\end{example}

Theorem~\ref{theo:ma_equivalences} demonstrates that  full monad algebra (say,
${\cal  M}_\cup[\sigma]$)  is  a  very  robust  notion.  It  can  serve  as
an ``expressiveness benchmark'' for query languages on complex-value
databases.   Indeed, it  has been  shown  that full  monad algebra  is a  {\em
conservative extension}\/ of relational algebra:\footnote{A generalized
version of Theorem~\ref{theo:conservative} can be found in \cite{Wong1996}.}

\begin{theorem}[\cite{PG1988}]
\label{theo:conservative}
A mapping from a (flat) relational database to a (flat) relation is
expressible in ${\cal M}_\cup[\sigma]$ if and only if it is expressible
in relational algebra.
\end{theorem}

It was discovered (see e.g.\ \cite{TBW1992} for a detailed discussion)
that virtually all the
complex value query languages developed in the eighties and nineties
that were intended for practical use are expressively equivalent.
Monad algebra is one of them, and so are
{\em nested relational
algebra}\/ \cite{JS82} and {\em complex value algebra without the powerset
operation}\/
\cite{AB1988,AHV95}.
Thus, the expressiveness results obtained in this paper for VCP
show that this language is equivalent to all of them.

\nop{
(Complex value algebra {\em with powerset}\/ \cite{AHV95} can
take hyperexponential runtime.
%
% $O(2^{2^{\dots 2^n}})$ in the size $n$ of the data, i.e.,
%proportional  to an  arbitrary tower  of twos.
%
Queries that  really  need the
powerset operator are usually too costly to evaluate.)
} % end nop

\section{The VCP Language}
\label{sect:vcp}

In this section, we introduce  VCP, a query language that
transforms complex value databases through a sequence of operations on a
schema tree.
VCP has two dynamic aspects, the first being how VCP operations transform the
schema tree at the time of specification (discussed in
Section~\ref{subsect:syntax}), and the second being the semantics of VCP
queries on data (Section~\ref{subsect:semantics}).
We provide three larger examples in Section~\ref{subsect:vcp_examples}.

\subsection{Query Specification as a Process}
\label{subsect:syntax}

Let $\mbox{nca}(v, e)$
denote the nearest common ancestor node of
node $v$ and edge $e$ in the schema tree.

The operations of the VCP language are presented in
Table~\ref{tab:operations}.
Operations 8 and 9 are implemented visually by drag\&drop
in the schema tree. The other operations
are local to a node or edge in the schema tree.

We have been parsimonious with the number of operations
that have been introduced in Table~\ref{tab:operations}; to make
query specification faster and more convenient,
further visual operations can be added, such as a
``delete subtree'' operation on set edges (which empties sets).
\footnote{This operation is redundant with
tuple insertion, subtree deletion on tuple edges, addition of constants, and
tuple elimination.}

\subsection{Informal Semantics}
\label{subsect:semantics}

\begin{figure*}[t!]
\begin{center}
\input{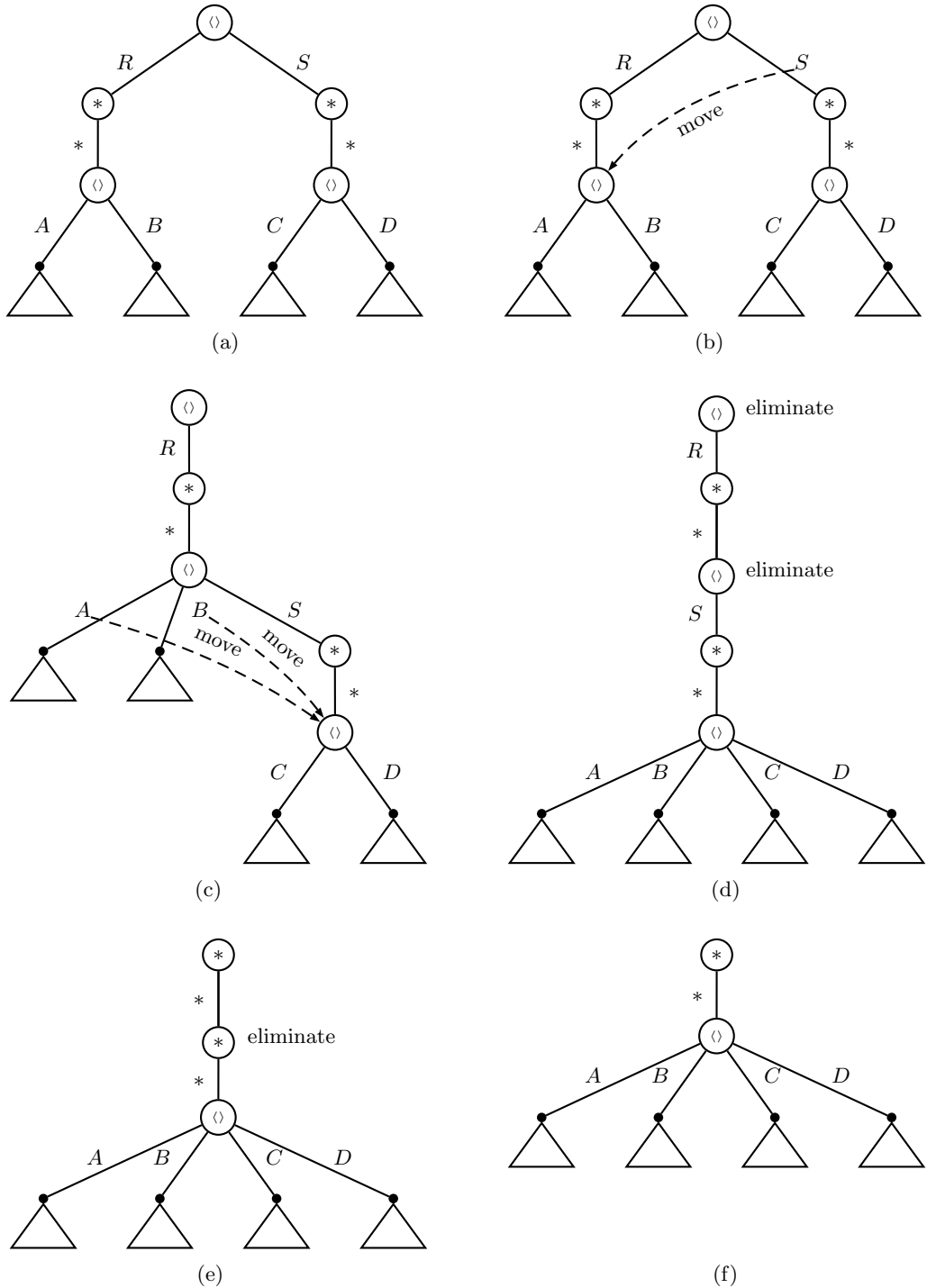}
\end{center}
\vspace{-5mm}
\caption{Simulating the cartesian product $R \times S$
of flat relational algebra in VCP.}
\label{fig:cartprod}
\end{figure*}

\begin{figure*}[t!]
\begin{center}
\input{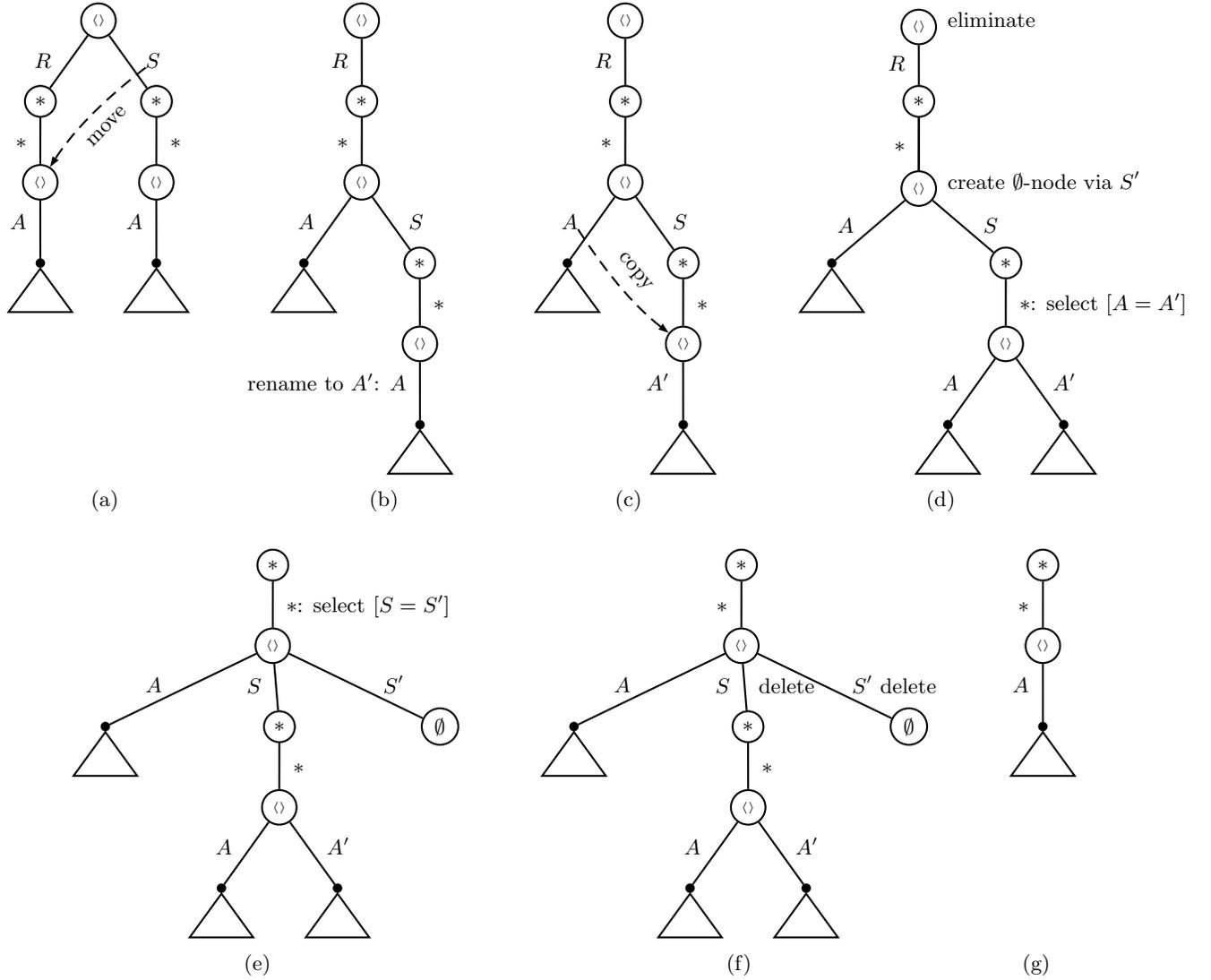}
\end{center}
\vspace{-4mm}
\caption{Modeling difference $R-S$ using selection in VCP.}
\label{fig:minus}
\end{figure*}

\begin{figure*}[t!]
\begin{center}
\input{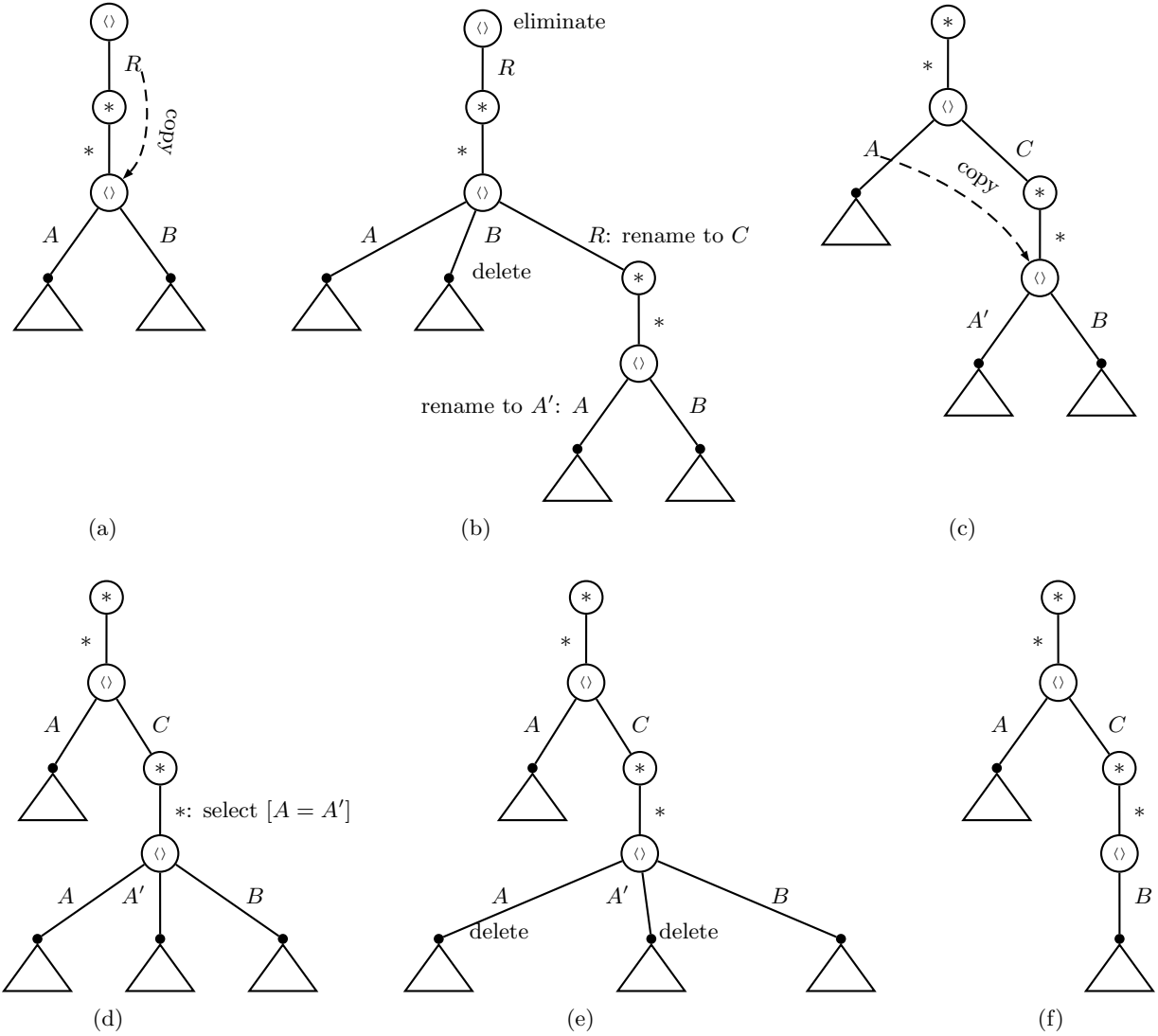}
\end{center}
\vspace{-4mm}
\caption{Nest$_{C=(B)}(R)$ in VCP.}
\label{fig:nest}
\end{figure*}

In order to obtain a simple informal semantics for VCP,
it is convenient to think of complex data values as trees themselves.
There are two main differences between schema trees and data trees. In the
former, $*$-nodes have precisely one child, while in the latter, $*$-nodes may
have many (one for each of the members of the set). Moreover,  leaf nodes
of schema trees are either labeled
``Dom'', $\emptyset$, or $\tuple{}$, while data
trees have atomic values instead of ``Dom'' at the leaves.
(See Figure~\ref{fig:ex1a}~{a} or \ref{fig:ex1b}~{a} for an example
of a schema tree and Figure~\ref{fig:ex1a}~{b} resp.\ \ref{fig:ex1b}~{b}
for a data tree compatible to the schema tree.
In order to avoid confusion -- but also to be economic with space --
all data trees in this paper are turned by 90 degreed compared
to schema trees.)

By the path between two nodes $v, w$ (where $v$ is an ancestor of $w$), we
denote the sequence of edge labels through which $w$ is reachable from $v$.
Note that in a schema tree, for any node $v$, each path $\pi$
uniquely identifies a node reachable from $v$ via $\pi$.
In other words, schema trees are deterministic trees. We can identify nodes of
the schema tree with their paths from the root node (e.g., we can talk of
``the node books.$*$.isbn'' in Figure~\ref{fig:ex1a}~(a)).
This is generally not
the case for data trees. A given path in the data tree may match a set of
nodes, e.g., relative to the root node of Figure~\ref{fig:ex1a}~(b),
books.$*$.year matches ``1988'' and ``2002''.

In the following, we say that a VCP operation is {\em schema-local}\/ to a
subtree of the schema tree if all the visual steps necessary to specify the
operation can be exclusively taken in the subtree.

There are three kinds of operations in VCP,
(a) operations on nodes (1, 2, 3, 5, 6, and 10
of Table~\ref{tab:operations}),
(b) operations on edges (4, 7, and 8
of Table~\ref{tab:operations}), and 
(c) copy-paste operations (8 and 9 of Table~\ref{tab:operations}).

\def\ctx{\mbox{ctx}}

By the {\em context node}\/ of an operation $o$, denoted $\ctx(o)$,
we refer to
\begin{itemize}
\item
the node of the schema tree that $o$ is applied to if $o$ is a
node operation (a), with the exception of {\em set note elimination}\/
where $\ctx(o)$ is its parent (a $*$-node as well),

\item
%the node the edge $o$ is applied to emanates from
the node that the edge emanates from which $o$ is applied to
if $o$ is an edge operation (b), and

\item
the node $\mbox{nca}(w, e)$ for a 
copy-operation (c) that copies from edge $e$ to node $w$.
\end{itemize}
Of course, $o$ is schema-local w.r.t.\ the subtree of $\ctx(o)$.

The {\em local semantics}\/ function
$L \Bracks{o}$ maps from a complex value to a complex value, or in
other words, from a data tree to a data tree:
$L \Bracks{\mbox{select(A=B)}}(u)$, where $u$ is a $*$-node
of the data tree,
removes all those members of set $u$ (= children of $u$ with their subtrees)
that do not satisfy the selection condition $A=B$.
$L \Bracks{\mbox{eliminate}}(u)$  flattens the set of sets $u$, i.e.,
replaces the subtree rooted by $u$ by $\bigcup u$.
For all other operations $o$ of types (a) and (b),
$L \Bracks{o}$ on $u$ is basically the same operation as described
in Table~\ref{tab:operations} for the schema tree applied to node $u$
(or an emanating edge) in the 
data tree.

The local semantics of copy-operations (c) is as follows.
For a copy-paste operation $o$ (type 8 or 9 in Table~\ref{tab:operations})
from source edge $e$ to destination node $w$, let
$\pi_{from}$ be the path from $\ctx(o)=\mbox{nca}(w,e)$
to $e$ and let $\pi_{to}$ be the destination path
from $\ctx(o)$ to $w$.
If $\pi_{to}$
contains $*$-edges, we say that $o$ is a bulk-copy operation.
If $o$ is on tuples (type 8),
$\pi_{from}$ consists only of tuple-nodes.
$L \Bracks{o}(u)$ is obtained from $u$ by copying
the source $u.\pi_{from}$ to each of the
tuple nodes matching $u.\pi_{to}$.
If $o$ is on sets (type 9),
$\pi_{from}$ consists only of tuple-nodes
and one set-node $s$ (the node which $e$ directly emanates from).
$L \Bracks{o}(u)$ is obtained from $u$
by copying each of the members of the source-set reachable from $u$ via path
$\pi_{from}$ (excluding the final ``$*$'') to
each of the destination sets matching $u.\pi_{to}$.

If the path $\pi$ from the root of the schema tree
to $ctx(o)$ contains $*$-nodes, we call $o$
a {\em bulk operation}\/.
Operation $o$ is executed by replacing
each node $u$ reachable through $\pi$ from the root of the data tree
by $L \Bracks{o}(u)$.

\begin{example}
{\em
Consider again the schema and the data tree of Figure~\ref{fig:ex1a}~(a) and
(b), respectively, 
The operation ``insert tuple (A)'' on the schema tree node
books.$*$.year  replaces the values of year attributes
in tuples reachable
through path books.$*$ in the data tree by a unary tuple node with the year
as the value of attribute $A$. 
For an edge manipulation example,
``rename to publyear'' on the schema tree node
books.$*$.year renames each of the ``year'' edges of tuple nodes
reachable through path books.$*$ of the data tree to ``publyear''.
}
\punto
\end{example}

\subsection{Examples}
\label{subsect:vcp_examples}

Figure~\ref{fig:cartprod} shows how the cartesian product $R \times S$
for two binary relations $R(A,B)$, $S(C,D)$ can be encoded in VCP.
Note that we employ two bulk copy operations (or, here, ``move'' to safe space)
to pair the tuples of the two relations.

The operation ``move'' is obtained
by first copying and subsequently deleting the source.
It is employed here to save space but with considerable headache;
a ``bulk copy'' operation is really much more intuitive than a
``move to many places'' operation.

In Figure~\ref{fig:minus}, we show how difference $R - S$ for unary relations
$R(A)$ and $S(A)$ can be encoded in VCP. The idea of the mapping is the same
as in Example~\ref{ex:m_minus}, but we have gone to the full length of
not assuming a selection operation of the form $\sigma_{A = \emptyset}$.
We use only a single form of selections, $\sigma_{A=B}$.
This query may also serve as a template for proving that
adding ``-'' to VCP does not extend its expressive power.

Of course, difference ``-'' could also be directly supported in a GUI based on
VCP to allow to realize this query in a single step. Difference ``$-$''
may not have
the clean visual metaphor of the other VCP operations, but it may be available
as an option for expert users.

Figure~\ref{fig:nest} shows how to implement the nest operation of complex
value algebra without powerset \cite{AB1988,AHV95} in VCP with selection.
We encode $\mbox{nest}_{C=(B)}(R)$ on relation $R(AB)$.

It it important to assert that each VCP query step is assumed to consist of a
single operation, even though we have sometimes resorted to annotating schema
trees in Figure~\ref{fig:cartprod}, \ref{fig:minus}, and \ref{fig:nest}
with several independent operations to safe space.

\section{Equivalence of VCP and Monad Algebra}
\label{sect:expressiveness}

In this section, we characterize the expressive power of VCP by showing that
it captures full monad algebra ${\cal M}_\cup[\sigma]$. We also show that
VCP without selection captures precisely positive monad algebra and that
VCP without the copy operation on sets and without selection
coincides with ${\cal
  M}$. These expressiveness results not only show that VCP captures the
``benchmark expressiveness'' for complex-value databases. The proofs of
the direction $VCP \subseteq {\cal M}_\cup[\sigma]$ also provide us with a
translation from VCP to full monad algebra that allows us to use existing
query engines and results on query evaluation for the latter language.
The proofs are straightforward and provide us with an alternative formal
semantics for VCP (through the monad algebra framework).

\begin{theorem}
VCP[wo `select'] $\subseteq {\cal M}_\cup$.
\end{theorem}

\noindent {\bf Proof Sketch}:
The proof is by induction. We define a function $\vcptom{\dots}$ that maps any
VCP query to an equivalent ${\cal M}$-expression.
\begin{itemize}
\item
If operation $o$ on schema
$\tuple{A_1: \tau_1, \dots, A_n: \tau_n}$
is schema-local to $\tau_1$, then
%
%\begin{multline*}
$
\vcptom{o} =
\langle A_1: \pi_{A_1} \circ \vcptom{o@A_1}, %\\
A_2: \pi_{A_2}, \dots, A_n: \pi_{A_n} \rangle
$
%\end{multline*}
%
where $o@A_1$ denotes $o$ modified to apply to the $A_1$ branch of the schema
tree (i.e., the subtree corresponding to $\tau_1$).

If operation $o$ on schema
$\tau = \{ \tau' \}$
is schema-local to $\tau'$, then
$
\vcptom{o} =
\mbox{map}(\vcptom{o@*})
$
where $o@*$ denotes the operation $o$ modified to apply to the subtree
corresponding to $\tau'$. 

For example, if $o$ is an operation schema-local to the subtree rooted
by node $v$ in the schema tree
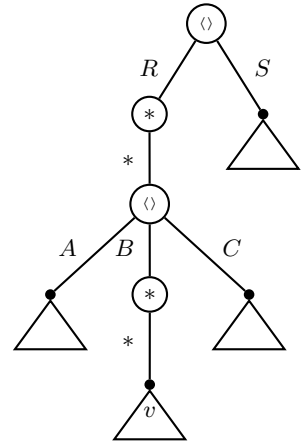
\begin{center}
\pstree[treefit=tight]{\Ttup}{
   \pstree{\Tset^{$R$}}{
      \pstree{\Ttup^{$*$}}{
         \Tsubtreel{$A$}
         \pstree{\Tset^{$B$}}{
            \pstree[levelsep=7.3mm]{\Tdot~{$v$}^{$*$}}{\Tfan}
         }
         \Tsubtreer{$C$}
      }
   }
   \Tsubtreer{$S$}
}
%
%\hspace{1cm}
%
%\pstree[treefit=tight]{\Ttup}{
%   \pstree{\TR{$\pi_R \circ \mbox{map}$}^{$R$}}{
%      \pstree{\Ttup}{
%         \TR{$\pi_A$}^{$A$}
%         \pstree{\TR{$\pi_B \circ \mbox{map}$}^{$B$}}
%         {
%            \TR{$\vcptom{(((o@R)@*)@B)@*}$}
%         }
%         \TR{$\pi_C$}_{$C$}
%      }
%   }
%   \TR{$\pi_S$}_{$S$}
%}
\end{center}
then
\begin{multline*}
\vcptom{o} =
\langle R:\pi_R \circ \mbox{map}(\langle A:\pi_A, \\
   B:\pi_B \circ \mbox{map}(\vcptom{(((o@R)@*)@B)@*}), \\
   C \circ \pi_C \rangle), S: \pi_S
\rangle.
\end{multline*}
%The close correspondence between schema trees and the term structure
%of ${\cal M}$-expressions obtained in this way is shown in
%Figure~\ref{fig:slocal} (right).

\item
The deletion of subtrees, the renaming of
tuple-edges, and the addition of new constants to tuples
is handled by tuple creation, constants, and projection.

\item
Tuple-node and set-node elimination is encoded using projection and
``flatten'', respectively:
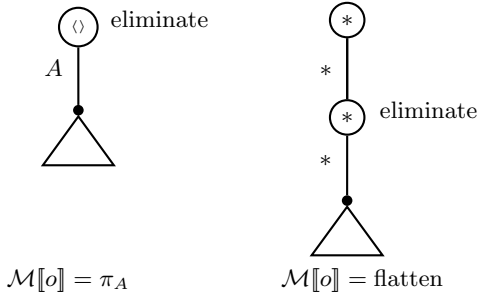
\begin{center}
   \pstree{\TR{{\white eliminate} $\quad$ \Ttup $\quad$ {eliminate}}}
                    {\Tsubtreel{$A$}}
%
%\hspace{5mm}
%
\pstree{\Tset}{
   \pstree{\TR{{\white eliminate} $\quad$ \Tset $\quad$ {eliminate}}^{$*$}}
                    {\Tsubtreel{$*$}}
}

~~ $\vcptom{o} = \pi_A$ \hspace{1.8cm}
$\vcptom{o} = \mbox{flatten}$
\end{center}

\item
A tuple-node with edge ``A'' is inserted by $\tuple{A: \mbox{id}}$.

\item
A set-node is inserted by ``$\mbox{sing}$''.

\item
Copy-paste expressions on tuples are mapped to ${\cal M}_\cup$ as shown
for two important cases:
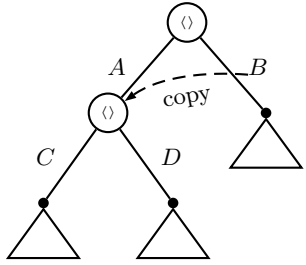
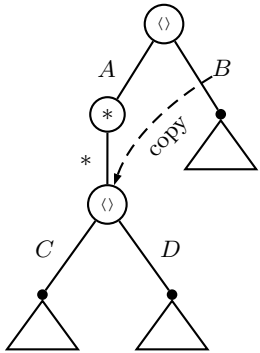
\begin{center}
\pstree[treefit=tight]{\Ttup}{
      \pstree{\Ttupn{To}^{$A$}}{
         \Tsubtreel{$C$}
         \Tsubtreer{$D$}
      }
   \Tsubtreer{\pnode{From}$B$}
}
\vcpl{From}{To}{20}
\end{center}
\[
\vcptom{o} =
\tuple{
A\!:\tuple{B\!: \pi_B, C\!:\pi_A \circ \pi_C, D\!: \pi_A \circ \pi_D},
B\!: \pi_B}
\]

\begin{center}
\pstree[treefit=tight]{\Ttup}{
   \pstree{\Tset^{$A$}}{
      \pstree{\Ttupn{To}^{$*$}}{
         \Tsubtreel{$C$}
         \Tsubtreer{$D$}
      }
   }
   \Tsubtreer{\pnode{From}$B$}
}
\vcpl{From}{To}{20}
\end{center}
\begin{multline*}
\vcptom{o} = \langle
A: \mbox{pairwith}_1(\pi_A, \pi_B) \circ
\mbox{map}(\langle B: \pi_2, \\
C:\pi_1 \circ \pi_C, D: \pi_1 \circ \pi_D
\rangle), B: \pi_B
\rangle
\end{multline*}

In general, the path from $\mbox{nca}(v,e)$ to $e$ can consist of 
of sequence of tuple edges,
from which be can extract the source node by a sequence
of projections.
On the path from $\mbox{nca}(v,e)$ to destination node $v$, there can be a
sequence of tuple nodes (handled using tuple construction as shown in the
first example above) and $*$-nodes,
handled using ``pairwith'' as in the second example.

\item
Copy-paste expressions on sets are mapped to union. For example,
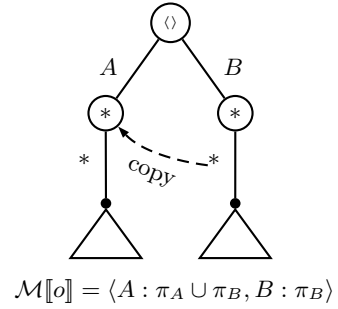
\begin{center}
\pstree[treefit=tight]{\Ttup}{
   \pstree{\Tsetn{To}^{$A$}}{
      \Tsubtreel{$*$}
   }
   \pstree{\Tset_{$B$}}{
      \Tsubtreel{\pnode{From}$*$}
   }
}
\vcpl{From}{To}{-20}

$\vcptom{o} = \tuple{A: \pi_A \cup \pi_B, B: \pi_B}$
\end{center}

\item
A sequence of VCP operations $O_1, \dots, On$ where each $O_i$
corresponds to an ${\cal M}$-expression $f_i$ simply corresponds to
composition $((f_1 \circ f_2) \circ \dots ) \circ f_n$.
\punto
\end{itemize}

The previous translation also yields

\begin{corollary}
VCP[wo `copy to set node', `select'] $\subseteq {\cal M}$.
\end{corollary}

For the other direction,

\begin{theorem}
${\cal M}_\cup \subseteq$ VCP[wo `select'].
\end{theorem}

\newcommand{\mtovcp}[1]{VCP \Bracks{#1}}

\noindent
{\bf Proof Sketch}:
The proof employs an inductive definition of a function  $VCP \Bracks{\dots}$
that maps ${\cal M}_\cup$ expressions to VCP queries.
Translating most operations is obvious, so we will just discuss the
most interesting:
\begin{itemize}
\item
map
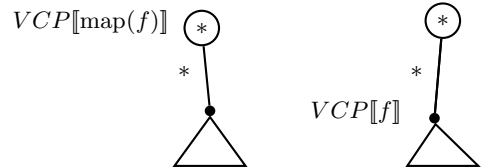
\begin{center}
\pstree{\TRtxt{$\mtovcp{\mbox{map}(f)}$}{\Tset}{~~~~}}{\Tsubtreel{$*$}}
\hspace{-1.5cm}
\pstree{\Tset}{
   \pstree[levelsep=7.3mm]{\TRtxt{$\mtovcp{f}$}{\Tdot}{~~~~}^{$*$}}{\Tfan}}
\end{center}

\item
pairwith. The essential operation here is copy-paste.
We implement $\mbox{pairwith}_{A_1}$ in VCP as follows.

\hspace{-2.2cm}
\pstree[treefit=tight]{\Ttup}{
   \pstree{\Tset^{$A_1$}}{
      \pstree[levelsep=7.3mm]{\TRtxtr{insert tuple($A_1$)}{\Tdot}}{\Tfan}
   }
   \Tsubtreel{$A_2$}
   \Tfan~{\dots}
   \Tsubtreer{$A_n$}
}
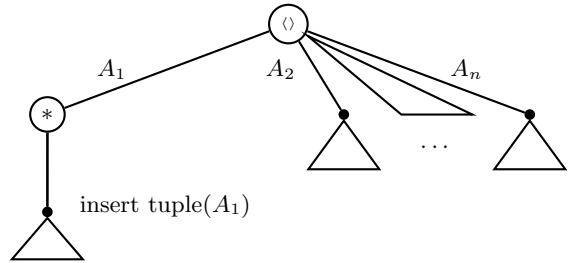
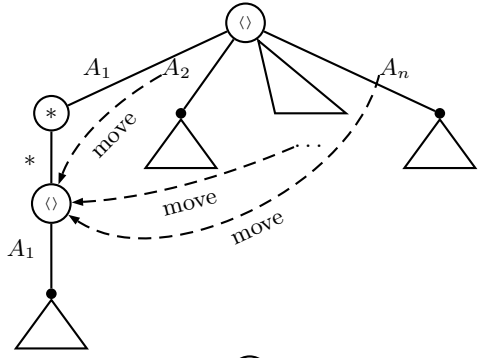
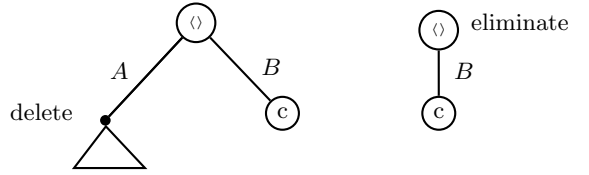
\begin{center}
\pstree{\Ttup}{
   \pstree{\Tset^{$A_1$}}{
      \pstree{\Ttupn{To}^{$*$}}{\Tsubtreel{$A_1$}}
   }
   \Tsubtreel{\pnode{From1}$A_2$}
   \Tfan~{\pnode{Fromi}\dots}
   \Tsubtreer{\pnode{Fromn}$A_n$}
}
\vcpmovel{From1}{To}{20}
\vcpmovel{Fromi}{To}{-10}
\vcpmovel{Fromn}{To}{-55}

\pstree{\TRtxtr{eliminate}{\Ttup}}{
   \pstree{\Tset^{$A_1$}}{
      \pstree{\Ttup^{$*$}}{
         \Tsubtreel{$A_1$}
         \Tsubtreel{$A_2$}
         \Tfan~{\dots}
         \Tsubtreer{$A_n$}
      }
   }
}
\end{center}

\item
tuple creation.
We rewrite each ${\cal M}$-expression of the form
$
\tuple{A_1: f_1, \dots, A_n: f_n}
$
into
\begin{multline*}
\tuple{A_1: \mbox{id}, \dots, A_n: \mbox{id}} \circ \\
\tuple{A_1: \pi_{A_1} \circ f_1, \dots, A_n: \pi_{A_n} \circ f_n}.
\end{multline*}
Now,
$\tuple{A_1: \mbox{id}, \dots, A_n: \mbox{id}}$ simply means to make a tuple
of $n$ copies of the input value which can be realized in VCP as
shown in Figure~\ref{fig:dupl} (for $n=3$).

The second step,
$\tuple{A_1: \pi_{A_1} \circ f_1, \dots, A_n: \pi_{A_n} \circ f_n}$,
just requires to push the $f_i$ down into the $A_i$ branches, for each
$1 \le i \le n$.
That is, we transform
\begin{center}
$VCP\lBrack \tuple{A_1: \pi_{A_1} \circ f_1, \dots, A_n:
	\pi_{A_n} \circ f_n} \rBrack$

%\pstree{\TRtxt{$VCP\lBrack \tuple{A_1: \pi_{A_1} \circ f_1, \dots, A_n:
%	\pi_{A_n} \circ f_n} \rBrack$}{\Ttup}{~~~~}}{
\pstree{\Ttup}{
   \Tsubtreel{$A_1$}
   \Tfan~{\dots}
   \Tsubtreer{$A_n$}
}
\end{center}
into

%\begin{center}
\hspace{-1cm}
\pstree[treefit=tight]{\Ttup}{
   \pstree[levelsep=7.3mm]{\TRtxt{$VCP \lBrack f_1 \rBrack$}{\Tdot}{~~}^{$A_1$}}{\Tfan}
   \Tfan~{\dots}
   \pstree[levelsep=7.3mm]{\TRtxt{$VCP \lBrack f_n \rBrack$}{\Tdot}{~~~~}_{$A_n$}}{\Tfan}
}
%\end{center}
%

\item
union is implemented using the copy operation between sets.

\item
constants.
\begin{center}
\pstree[levelsep=7.3mm]{\TR{$\stackrel{\mbox{insert tuple($A$)}}{\Tdot}$}}{\Tfan}
\hspace{1cm}
\pstree{\TR{$\begin{array}{c}
\mbox{add constant $c$ via $B$} \\
\mbox{\Ttup}
\end{array}$}}{
   \Tsubtreel{$A$}
}

\pstree[treefit=tight]{\Ttup}{
   \pstree[levelsep=7.3mm]{\TRtxt{delete}{\Tdot}{$\quad$}^{$A$}}{\Tfan}
   \Tcircle{c}_{$B$}
}
\pstree{\TRtxtr{eliminate}{\Ttup}}{\Tcircle{c}_{$B$}}
\end{center}
\punto
\end{itemize}

\begin{corollary}
${\cal M} \subseteq$ VCP[wo `copy to set node', `select'].
\end{corollary}

The same selection operation is available in VCP and ${\cal M}_\cup[\sigma]$.
From the previous results, we obtain

\begin{theorem}
$VCP = {\cal M}_\cup[\sigma]$.
\end{theorem}

%%%%%%%%%%%%%%%%%%%%%%%%%%%%%

\begin{figure*}[t]
\begin{center}
\input{fig_dupl.tex}
\end{center}
%\vspace{-2mm}
\caption{{\rm $VCP
\Bracks{\tuple{A_1: \mbox{id}, A_2: \mbox{id}, A_3: \mbox{id}}}$.}}
\label{fig:dupl}
\end{figure*}

\section{Discussion and Conclusions}
\label{sect:discussion}

In our presentation, we have implicitly assumed set-typed collections,
but at least for positive monad algebra ${\cal M}_\cup$ and for
VCP[wo selection], sets can 
be replaced by bags or
lists and we immediately obtain that ${\cal M}_\cup = VCP$[wo selection]
still holds.
For the practical viewpoint, this implies that VCP can be used for visually
specifying
bag queries (as in SQL), but it also exhibits one oblique notion
that users have to understand in order to successfully use the VCP language,
namely that of collections. Under the set-interpretation, dragging a
$*$-edge onto a $*$-node results in the addition of the members of the source
sets to the destination set {\em with duplicate elimination}\/, while for bags,
duplicates are not eliminated, and for lists, this operation means to append
the source list to the destination list.

Regarding full monad algebra, there is
unfortunately no agreed-upon notion of intersection or
difference for bags and lists (several alternatives can be
reasonably justified). Moreover, not all of these notions lead to the
same expressive power when added to ${\cal M}_\cup$.
For example, the probably most natural notion 
of difference on bags usually referred to as {\em monus}\/ is known to yield
the power of arithmetics, while this is not the case for intersection or
selection \cite{LW1997}.

Still, any of these differing ``full'' bag or list monad algebras can
be simulated in VCP by just adding the desired operations (such as selection
or difference) directly. Of course this is a pragmatic
solution, but actually not more daring than what we have done in the set case
earlier in the paper: positive monad algebra was realized using just
insertion, renaming, deletion, and copying, while
the move to the expressiveness of full monad algebra
was made by adding a selection
operation.
The choice of which operations are
made available in the language (in the case of sets, selection, intersection,
and difference are interchangeable and even adding them all does not yield
greater expressiveness)
should depend on  which operations appear natural given the 
design choices made for the GUI of the query editor.

One future area of research will be to provide a visual language for defining
XML queries on the basis of VCP. Ordered, unranked XML trees can
be viewed as nested lists, but to get a visual language for XML with the
expressive power of (full) monad algebra on lists, also tuple-typed nodes
are required. These do not exist in XML, but could be simulated using
the XPath position() function; that is, the children of pseudo-tuple nodes
in XML would only be accessed using paths of the form ``child[position() =
  $i$]'', yielding the $i$-the attribute of the pseudo-tuple.

Schema information for XML is usually provided in the form of Document Type
Definitions or XML Schemas, which correspond to possibly infinite schema trees
(through recursion). However, this is not a problem in VCP or VCP extensions
that correspond to XQueries that use no transitive axes. In VCP, each query
only navigates into a schema tree up to a certain depth fixed with the query.
Further work will be required for VCP to deal with navigation to the
descendants of XML nodes.

VCP -- in an extended form -- seems to be a promising candidate for
a bridging formalism
between (relational) databases, file systems, and (XML or HTML) data
trees. Such a bridge could lead to greatly enhanced usability of visual tools
for Web site and Web service definition.
This is one possible direction of future research.

\nop{
\section*{Acknowledgments}

I am indebted to Peter Buneman for many insightful discussions on data
provenance and query languages. These have inseminated the present
work.
} % end nop

%\begin{todo}
%Acknowledge CP \cite{BK}.
%\end{todo}

\bibliographystyle{abbrv}
{\small
\bibliography{bibtex}

\newcommand{\SortNoOp}[1]{}
\begin{thebibliography}{10}

\bibitem{AB1988}
S.~Abiteboul and C.~Beeri.
\newblock {``The Power of Languages for the Manipulation of Complex Values''}.
\newblock {\em VLDB J.}, {\bf 4}(4):727--794, 1995.

\bibitem{AHV95}
S.~Abiteboul, R.~Hull, and V.~Vianu.
\newblock {\em {Foundations of Databases}}.
\newblock Addison-Wesley, 1995.

\bibitem{XQBE}
E.~Augurusa, D.~Braga, A.~Campi, and S.~Ceri.
\newblock {``Design and Implementation of a Graphical Interface to XQuery''}.
\newblock In {\em Proc.\ SAC}, pages 1163--1167, Melbourne, Florida, 2003.

\bibitem{BFG2001b}
R.~Baumgartner, S.~Flesca, and G.~Gottlob.
\newblock {``Declarative Information Extraction, Web Crawling, and Recursive
  Wrapping with Lixto''}.
\newblock In {\em Proc. LPNMR'01}, Vienna, Austria, 2001.

\bibitem{BFG2001a}
R.~Baumgartner, S.~Flesca, and G.~Gottlob.
\newblock {``Visual Web Information Extraction with Lixto''}.
\newblock In {\em Proceedings of the 27th International Conference on Very
  Large Data Bases (VLDB)}, pages 119--128, Rome, Italy, 2001.

\bibitem{BBSW2003}
S.~Berger, F.~Bry, S.~Schaffert, and C.~Wieser.
\newblock {``Xcerpt and visXcerpt: From Pattern-Based to Visual Querying of XML
  and Semistructured Data''}.
\newblock In {\em Proc.\ VLDB (Demo Track)}, pages 1053--1056, 2003.

\bibitem{BNTW1995}
P.~Buneman, S.~A. Naqvi, V.~Tannen, and L.~Wong.
\newblock {``Principles of Programming with Complex Objects and Collection
  Types''}.
\newblock {\em Theor.\ Comput.\ Sci.}, {\bf 149}(1):3--48, 1995.

\bibitem{XMLGL}
S.~Comai, E.~Damiani, and P.~Fraternali.
\newblock {``Computing Graphical Queries over XML Data''}.
\newblock {\em ACM TOIS}, {\bf 19}(4):371--430, 2003.

\bibitem{CoMe90}
M.~P. Consens and A.~O. Mendelzon.
\newblock {``GraphLog: a Visual Formalism for Real Life Recursion''}.
\newblock In {\em Proceedings of the 9th ACM SIGACT-SIGMOD-SIGART Symposium on
  Principles of Database Systems (PODS'90)}, 1990.

\bibitem{CMW1988}
I.~F. Cruz, A.~O. Mendelzon, and P.~T. Wood.
\newblock {``G+: Recursive Queries Without Recursion''}.
\newblock In {\em Expert Database Conf.}, pages 645--666, 1988.

\bibitem{GKJACM}
G.~Gottlob and C.~Koch.
\newblock {``Monadic Datalog and the Expressive Power of Web Information
  Extraction Languages''}.
\newblock {\em Journal of the ACM}, {\bf 51}(1):74--113, 2004.

\bibitem{JS82}
G.~Jaeschke and H.-J. Schek.
\newblock {``Remarks on the Algebra of Non First Normal Form Relations''}.
\newblock In {\em Proc. PODS'82}, pages 124--138, 1982.

\bibitem{LW1997}
L.~Libkin and L.~Wong.
\newblock {``Query Languages for Bags and Aggregate Functions''}.
\newblock {\em J.\ Comput.\ Syst.\ Sci.}, {\bf 55}(2):241--272, 1997.

\bibitem{PPV2002}
Y.~Papakonstantinou, M.~Petropoulos, and V.~Vassalos.
\newblock {``QURSED: Querying and Reporting Semistructured Data''}.
\newblock In {\em Proc.\ SIGMOD Conference}, pages 192--203, 2002.

\bibitem{PPT1995}
J.~Paredaens, P.~Peelman, and L.~Tanca.
\newblock {``G-Log: A Graph-Based Query Language''}.
\newblock {\em IEEE Trans.\ Knowl.\ Data Eng.}, {\bf 7}(3):436--453, 1995.

\bibitem{PG1988}
J.~Paredaens and D.~{Van Gucht}.
\newblock {``Possibilities and Limitations of Using Flat Operators in Nested
  Algebra Expressions''}.
\newblock In {\em Proc.\ PODS}, pages 29--38, 1988.

\bibitem{TBW1992}
V.~Tannen, P.~Buneman, and L.~Wong.
\newblock {``Naturally Embedded Query Languages''}.
\newblock In {\em Proc.\ ICDT}, pages 140--154, 1992.

\bibitem{Wong1996}
L.~Wong.
\newblock {``Normal Forms and Conservative Extension Properties for Query
  Languages over Collection Types''}.
\newblock {\em J.\ Comput.\ Syst.\ Sci.}, {\bf 52}(3):495--505, 1996.

\bibitem{QBE}
M.~M. Zloof.
\newblock {``Query-by-Example: A Data Base Language''}.
\newblock {\em IBM Systems Journal}, {\bf 16}(4):324--343, 1977.

\end{thebibliography}
}

\end{document}